\documentclass[11pt,a4paper]{article}
\usepackage{jheppub}
\usepackage[T1]{fontenc} 
\usepackage{comment}
\usepackage{scalerel}
\usepackage{upgreek}
\newcommand{\tmin}{T_{\scaleto{\rm min}{4.5pt}}}
\newcommand{\tc}{T_{\scaleto{\rm c}{3pt}}}
\newcommand{\fbh}{F_{\scaleto{\rm BB}{4pt}}}
\newcommand{\sbb}{s_{\scaleto{\rm BB}{4pt}}}
\newcommand{\ssbh}{s_{\scaleto{\rm SBB}{4pt}}}
\newcommand{\fsbh}{F_{\scaleto{\rm SBB}{3.7pt}}}
\newcommand{\fdads}{F_{\scaleto{\rm dAdS}{4.5pt}}}
\newcommand{\phih}{\phi_{\scaleto{\rm h}{4.5pt}}}
\newcommand{\phimin}{\phi_{\scaleto{\rm min}{4.5pt}}}
\newcommand{\phic}{\phi_{\scaleto{\rm c}{3pt}}}
\newcommand{\uh}{u_{\scaleto{\rm h}{4.5pt}}}
\newcommand{\mpld}{M_{\scaleto{D+1}{4.5pt}}}
\newcommand{\cs}{c_{\scaleto{\rm s}{3.5pt}}}
\newcommand{\ccft}{c_{\scaleto{\rm s}{3.5pt},\scaleto{\rm CFT}{3.5pt}}}
\newcommand{\epsdw}{\epsilon}

\author[\textsuperscript{\fontsize{10}{11}\selectfont$\phi$}]{Prateek Agrawal,}
\author[\textsuperscript{\fontsize{10}{11}\selectfont$\upvarphi$}]{Gaurang Ramakant Kane,}
\author[\textsuperscript{\fontsize{10}{11}\selectfont$\upvarphi$}]{and Vazha Loladze}
\title{\Large Localised Horizons and Holographic Thermodynamics: Supercooling in the 1/D Expansion}
\affiliation[\textsuperscript{\fontsize{10}{11}\selectfont$\phi$}]{Department of Physics, University of California, Santa Barbara, CA 93106, USA}
\affiliation[\textsuperscript{\fontsize{10}{11}\selectfont$\upvarphi$}]{Rudolf Peierls Centre for Theoretical Physics, University of Oxford, Parks Road, Oxford OX1 3PU, United Kingdom}
\emailAdd{prateekagrawal@ucsb.edu}
\emailAdd{gaurang.kane@physics.ox.ac.uk}
\emailAdd{vazha.loladze@physics.ox.ac.uk}
\abstract{
In holography, four-dimensional confining gauge theories are often modelled by five-dimensional Einstein--scalar gravity by choosing a specific form of the scalar potential. In a large class of non-conformal theories, we show that a predictive structure emerges for the thermal confinement transition by generalising the gravitational dual to $D+1$ dimensions and using a $1/D$ expansion. These results are independent of the details of the scalar potential, hinting towards universality. The black brane geometry dual to the deconfined phase can be analytically constructed due to its effects being localised near the horizon at leading order. The solution does not exist below a minimal temperature $\tmin$ and the maximum possible supercooling in the transition $\epsilon_{\rm sc} = 1-\tmin/T_{\rm c}$ is generically suppressed by a factor of $1/D^2$. 
Remarkably, the maximum supercooling at the leading order is set by the speed of sound in the deconfined phase of the gauge theory at the critical temperature, $\epsilon_{\rm sc}=c_s^2(\tc)/2$. These predictions agree with explicit calculations in an exponential superpotential, improved holography, and  the thermal transition in $\mathcal{N}=4$ super Yang--Mills on a sphere. 
}

\begin{document}
\maketitle
\section{Introduction}
\label{sec:introduction}
The understanding of the phases and dynamics of phase transitions in strongly coupled gauge theories remains one of the major unsolved problems in quantum field theory. The underlying physics has been explored using several complementary techniques, including lattice gauge theories \cite{Polyakov:1978vu, McLerran:1981pb, Kuti:1980oza, Yaffe:1982qf, Lucini:2002ku, Lucini:2003zr, Crean:2026caz}, supersymmetric constructions and their deformations \cite{Seiberg:1994rs,Seiberg:1994aj, Douglas:1995nw,Hanany:1997hr, Davies:2000nw, Aharony:2003sx, Poppitz:2012sw, Poppitz:2012nz, Anber:2017tug}, and holography \cite{Witten:1998zw,Gubser:1999pk, Klebanov:2000nc, Klebanov:2000hb, Arkani-Hamed:2000ijo, Rattazzi:2000hs, Creminelli:2001th,Karch:2006pv,Randall:2006py, Gursoy:2007cb, Gursoy:2007er,Hassanain:2007js,Batell:2008zm, Gubser:2008ny, Gursoy:2008za, Konstandin:2011dr, Agashe:2020lfz,Hanada:2022wcq, Mishra:2024ehr}.
These techniques have provided valuable insights into the properties of the confined and deconfined phases of the gauge theories and the dynamics of the thermal transitions between them. Nevertheless, many aspects of the problem remain poorly understood.
Confinement transitions may also have important phenomenological consequences (see \cite{Asadi:2026mip} for a recent review). There are many well-motivated beyond the Standard Model (BSM) scenarios in which a strongly coupled sector undergoes the confinement transition in the early Universe \cite{Kaplan:1983fs, Georgi:1984af, Kaplan:1991dc, Contino:2003ve, Agashe:2004rs, Arkani-Hamed:2001nha, Chacko:2005pe, Kribs:2016cew}. Such transitions can leave observable cosmological signatures and provide a rare window into strongly coupled BSM physics.

A powerful approach to studying confinement transitions is through the holographic principle. In these setups, the thermal deconfinement--confinement transition in the strongly coupled gauge theory \cite{Witten:1998zw} corresponds to the phase transition between a black brane geometry and the AdS geometry (the Hawking--Page transition \cite{Hawking:1982dh}) on the gravitational side. A broad class of phenomenologically motivated holographic models can be described as deformations of AdS space by a scalar field with a potential, including Randall--Sundrum models \cite{Randall:1999ee} and improved holographic Yang--Mills theory \cite{Gubser:1999pk,Gursoy:2007cb,Gursoy:2007er,Gursoy:2008za}. These constructions can also be used to model string-theoretic constructions such as the Klebanov--Tseytlin~\cite{Klebanov:2000nc} and the Klebanov--Strassler~\cite{Klebanov:2000hb} solutions. 

Studying the Hawking--Page transition in the gravitational theory requires solving the nonlinear Einstein--scalar equations for the black brane and the deformed AdS geometry. The case where the scalar field has a small backreaction on AdS geometry has been well-studied in Randall--Sundrum (RS) models with the simplest Goldberger--Wise stabilisation~\cite{Goldberger:1999uk, Creminelli:2001th, Randall:2006py, Konstandin:2011dr, Baratella:2018pxi, Agashe:2019lhy, Agashe:2020lfz} and captures near-conformal theories. Modelling non-conformal theories requires going beyond this approximation to a large scalar backreaction regime. One technique to solve the nonlinear equations at large backreaction is the superpotential technique introduced in \cite{Behrndt:1999kz,Skenderis:1999mm, Hatanaka:1999ac, DeWolfe:1999cp}, which can be used to find the deformed AdS geometry, but this does not work generically for the black brane solution~\cite{Dhumuntarao:2019eni}.
The available studies typically rely on numerical solutions, making it difficult to isolate the physics that determines the dynamics of the phase transition.

In this work, we construct black brane solutions as a perturbative expansion in $1/D$ based on studying gravity in a large number of spacetime dimensions $D \gg 1$~\cite{Emparan:2013moa,Emparan:2015hwa,Herzog:2017qwp,Emparan:2025yfy, Lee:2026uvo} (see \cite{Emparan:2020inr} for a comprehensive review). We can then extract some thermodynamic quantities which are insensitive to the details of the potential, and hence possibly universal. 

An important such quantity for the confining phase transition is the maximum possible supercooling, $\epsilon_{\rm sc}=1-T_{\rm min}/T_{\rm c}$, set by the minimal temperature of the deconfined phase $\tmin$ and the critical temperature, $T_{\rm c}$. For small supercooling, the phase transition can reheat the universe back up to critical temperature and the phase transition can be stalled, completely modifying its dynamics~\cite{Witten:1984rs, GarciaGarcia:2015fol, Gouttenoire:2023roe}. The amount of supercooling has dramatic consequences for the gravitational wave signals~\cite{Caprini:2019egz, Pasechnik:2023hwv, Huber:2025qbl}.

The degree of supercooling is closely tied to conformality of the theory near the transition. Strongly coupled gauge theories with approximate conformal invariance undergo significant supercooling (and can produce loud gravitational-wave signals). This has been studied using dilaton effective theories \cite{Creminelli:2001th, Randall:2006py, Konstandin:2011dr, Baratella:2018pxi, Agashe:2019lhy, Agashe:2020lfz} and by explicitly constructing strongly coupled gauge theories with near-conformal dynamics \cite{Miura:2018dsy, Azatov:2020nbe, Fujikura:2025iam, Agrawal:2025wvf}. 

In contrast, theories that are far from conformality near the transition appear to exhibit only a small amount of supercooling. On the lattice, results for large-$N$ $SU(N)$ gauge theories show that the phase transition near the critical temperature is very strongly first-order, although with a small numerical prefactor, which suggests that the maximum possible supercooling is small, $\epsilon_{\rm sc}  \ll 1$~\cite{Agrawal:2025xul}. This behaviour can also be seen in specific examples in holography where a large backreaction is included in the RS setup~\cite{Mishra:2024ehr,Ismail:2026tyb}, holographic models motivated by the Klebanov--Strassler solution \cite{Klebanov:2000hb,Klebanov:2000nc, Hassanain:2007js, Buchel:2009bh}, improved holographic QCD~\cite{Gursoy:2007cb,Gursoy:2007er,Gursoy:2008za, Morgante:2022zvc}, and soft-wall constructions \cite{Karch:2006pv,Batell:2008zm}. While these results are suggestive, they rely on specific assumptions of the scalar potential.

In our framework we obtain a generic prediction for the supercooling in confinement transitions in a large class of non-conformal theories.
We show that the maximum possible amount of supercooling is suppressed by a factor of $1/D^2$. The maximum supercooling is related to the speed of sound $\cs$ in the deconfined phase at the critical temperature. Within the regime of validity of our approximation, we find at leading order that $\epsilon_{\rm sc}=\cs^2(T_{\rm c})/2$. We verify that this relation is respected in explicit examples in section~\ref{sec:exponential_example}. We emphasise, however, that our result does not apply to theories with approximate conformal invariance near the confinement scale, where the assumptions underlying our analyses break down. 

\section{Einstein--scalar gravity in \texorpdfstring{$D+1$}{D+1} dimensions}
\label{sec:superpotential}
Five-dimensional Einstein gravity coupled to a scalar field on asymptotically AdS space is often used to model strongly coupled gauge theories in four dimensions through the holographic principle \cite{Gubser:1999pk, Klebanov:2000hb,Klebanov:2000nc, Arkani-Hamed:2000ijo,Rattazzi:2000hs, Creminelli:2001th, Karch:2006pv, Gursoy:2007cb, Gursoy:2007er,Batell:2008zm, Hassanain:2007js, Gubser:2008ny, Gursoy:2008za}. General solutions of the system are difficult to obtain, especially in the interesting regime where the backreaction of the scalar field on the background geometry is important. We show that the dynamics of a more general $(D+1)-$dimensional Einstein--scalar system can be analytically solved in a $1/D$ expansion for a large class of scalar potentials, including ones where the backreaction is important. The action of this theory can be written as
\begin{equation}
    S=-\dfrac{\mpld^{D-1}}{2}\int d^{D+1}x~\sqrt{\vert g\vert}~\left(R-\dfrac{1}{2}(\partial\phi)^{2}-V(\phi)\right)+S_{\rm GHY}~~,
    \label{eqn:einstein_scalar_action}
\end{equation}
where $S_{\rm GHY}$ is the Gibbons--Hawking--York boundary term. In the above convention, $\phi$ is a dimensionless field. 

We are interested in studying the thermal deconfinement--confinement transition in a strongly coupled gauge theory on flat space. Therefore, we use the following metric ansatz with a Euclidean metric which has flat $D-1$ spatial slices, a warped spatial dimension and a compact time direction with $t\equiv t+T^{-1}$, $T$ being the temperature
\begin{equation}
ds^{2}=e^{2A(u)}\left(f(u) dt^{2}+\delta_{ij}dx^{i}dx^{j}\right)+\dfrac{du^{2}}{f(u)}~~.
\label{eqn:warp_metric_ansatz}
\end{equation}
For studying translation-invariant stationary states, the metric and the scalar can only depend on the coordinate $u$ \cite{Bardeen:1973gs}. The equations of motion with this metric ansatz are:
\begin{align}
 \label{eqn:equation_of_motion}
 \ddot{A}
 + \dfrac{1}{2(D-1)}\dot{\phi}^{2}  &= 0,
 \\ \nonumber
 \dfrac{\ddot{f}}{\dot{f}}+D\dot{A}
 &= 0,
\\ \nonumber
f\ddot{\phi}+\dot{f}\dot{\phi}+Df\dot{A}\dot{\phi}
&=
\dfrac{dV}{d\phi},
\\
D(D-1)f\dot{A}^{2}+(D-1)\dot{A}\dot{f}&=\dfrac{f\dot{\phi}^{2}}{2}-V \,,
\nonumber
\end{align}
where the dot represents a derivative with respect to the coordinate $u$. Note that only three out of the four equations above are independent. A detailed derivation of these equations is presented in appendix \ref{appendix:details_of_the_equations_of_motion}. 

Let us first look at the effect of the $1/D$ expansion in the absence of a scalar field with just a negative cosmological constant included. In this case, the warp factor takes the form 
\begin{equation*}
A(u)=-\dfrac{u}{L}~~.
\end{equation*}
The $u\rightarrow -\infty$ limit corresponds to the asymptotic boundary and $L$ is the AdS curvature scale. Further, $f(u)$ is the blackening factor that captures the potential presence of a black brane horizon. Without the black brane the geometry is purely $\text{AdS}_{D+1}$ and $f(u)=1$. In the geometry with a black brane $f(u)$ goes to zero at the horizon. In this AdS--Schwarzschild black brane geometry, the blackening factor is
\begin{equation}
    f(u)=1-e^{D(u-\uh)/L}~~,
\label{eqn:pure_ads_blackening_factor}
\end{equation}
\begin{figure}[t!]
    \centering
\includegraphics[width=0.7\linewidth]{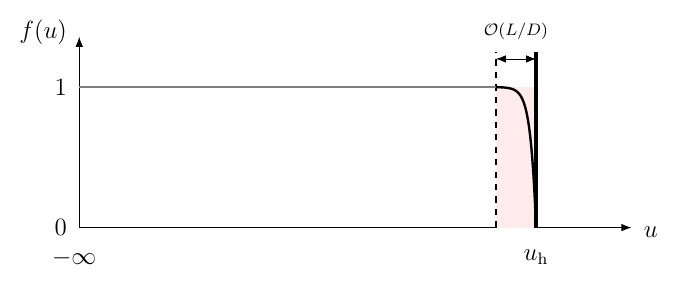}
    \caption{A schematic picture of the blackening factor for $\text{AdS}_{D+1}$-Schwarzschild geometry given in equation \eqref{eqn:pure_ads_blackening_factor}. Here $L$ is the AdS curvature scale. For large values of $D$, there is an effective region of order $L/D$ near the horizon given by $u=\uh$ where the blackening factor $f(u)$ changes rapidly.}
\label{fig:Schematic_icture_pure_ads}
\end{figure}
where $\uh$ is the position of the horizon in the bulk. Crucially, the blackening factor is effectively one for $\vert u-\uh\vert\gg L/D$. In the limit $D\gg 1$, we get an additional scale $L/D$ within which the black brane effects are localised and parametrically separated from the AdS scale $L$. This observation shall play an important role in section~\ref{sec:1_d_expansion}.

After including the scalar field and the potential, the pure AdS geometry gets deformed. Note that in holographic constructions, one is interested in deformations such that near the asymptotic boundary corresponding to $u\rightarrow -\infty$, the geometry is still AdS at leading order and the deformation becomes relevant as one moves deep in the bulk. We shall refer to such constructions as ``dAdS'' for short. It is useful to note that, by definition, $f(\uh)=0$ at the horizon. Hence, requiring the equations of motion \eqref{eqn:equation_of_motion} to remain regular at the horizon, we obtain:
\begin{equation}
    \dot{f}(\uh)=-\dfrac{V(\phih)}{(D-1)\dot{A}(\phih)}=\dfrac{1}{\dot{\phi}(\uh)}\dfrac{dV}{d\phi}(\phih)~~,
    \label{eqn:regularity_at_horizon}
\end{equation}
where $\phi(\uh)\equiv \phih$ is the value of the scalar field at the horizon.  

In the absence of the black brane, $f(u)=1$ throughout the entire spacetime. In this case, the independent equations of motion are:
\begin{align}
 &\ddot{A}=-\dfrac{1}{2(D-1)}\dot{\phi}^{2}~,~~D(D-1)\dot{A}^{2}=\dfrac{\dot{\phi}^{2}}{2}-V~~.
\end{align}
Although this is still a set of coupled nonlinear differential equations, as shown in~\cite{DeWolfe:1999cp, Hatanaka:1999ac, Behrndt:1999kz, Skenderis:1999mm}, it can be solved exactly when the potential $V(\phi)$ can be written in terms of a superpotential $W(\phi)$. In this case, the warp factor $A(u)$, the scalar field profile $\phi(u)$ and the potential $V(\phi)$ can be expressed in terms of $W(\phi)$ and its derivatives as
\begin{equation}
    \dot{A}=-W~,~~\dot{\phi}=2(D-1)\dfrac{dW}{d\phi}~,~
   V= 2(D-1)^{2}\left(\left(\dfrac{dW}{d\phi}\right)^{2}-\dfrac{D}{2(D-1)}W^{2}\right)~~.
   \label{eqn:vacuum_superpotential}
\end{equation}
As pointed out in \cite{Townsend:1984iu, Skenderis:2002wp}, based on the positive-energy argument, for a single scalar field in an asymptotically AdS spacetime, the potential can always be expressed in terms of a superpotential\footnote{In the case of multiple scalar fields, this is not necessarily true any more.}. Hence, the superpotential method can be used to obtain all the relevant vacuum solutions of our interest. However, to study the phase transition one must also construct black brane solutions. As shown in \cite{Dhumuntarao:2019eni}, in the presence of the black brane, the superpotential method is severely restricted by the condition $f(u_{\rm h})=0$ and can be used only for exponential superpotentials of the form $W=e^{\gamma \phi}$ also known as the Chamblin-Reall solution~\cite{Chamblin:1999ya}. Nevertheless, in the following section, we show that we can still study the thermodynamics of black branes for a large class of superpotentials by studying theories in a large number of spacetime dimensions. 

\section{The \texorpdfstring{$1/D$}{1/D} expansion}
\label{sec:1_d_expansion}
As discussed above, the dAdS solution can be obtained at large scalar backreaction by using the superpotential method, but this does not work for finding the black brane solution in general. Here, we overcome this obstacle by combining the superpotential method with the $1/D$ expansion. 

Let us start by defining the local AdS curvature by $L_{\rm eff}=\vert \dot{A}\vert^{-1}$. We are interested in gravitational solutions that are holographically dual to gauge theories using the AdS/CFT correspondence. Therefore, we take the geometry to approach AdS near the asymptotic boundary ($u\rightarrow-\infty$), implying $\left|dL_{\rm eff}/du\right|\ll 1$ sufficiently close to the boundary. Furthermore, to model a confining theory we take a relevant CFT deformation, so that $\left|dL_{\rm eff}/du\right|$ grows as one moves in the bulk.

The crucial observation we made in section \ref{sec:superpotential} is that the appearance of $D$ introduces a scale separation between the effective local AdS curvature scale and the shorter length scale $L_{\rm eff}/D$. As long as $\left|dL_{\rm eff}/du\right|\ll D$, the second equation of \eqref{eqn:equation_of_motion} implies that the blackening function is localised near the horizon: it changes from $f=0$ at the horizon to $f=1$ in the bulk over a distance scale of order $\vert L_{\rm eff}/D\vert$. This separates the bulk into near ($\vert u-\uh \vert \ll L_{\rm eff}$) and far horizon ($\vert u-\uh\vert \gg L_{\rm eff}/D$) regions. We can solve the equations of motion separately in these regions and since they overlap, match them where both solutions are valid.

The condition $\left|dL_{\rm eff}/du\right|\ll D$ is valid in a domain extending from the asymptotic boundary to the point in the bulk where the variation of the local AdS curvature becomes parametrically large. Note that the AdS isometries can be significantly broken while the approximation is still valid due to the $1/D$ parametrics. We expect precisely this region to contain the thermodynamic behaviour of the phase transition. In terms of the superpotential, the condition $\vert dL_{\rm eff}/du\vert\ll D$ is equivalent to
\begin{equation}
  \left(\dfrac{W'(\phi)}{W(\phi)}\right)^{2}\ll \dfrac{D}{2(D-1)}\,.
    \label{eqn:upper_bound_condition}
\end{equation}
Additionally, in order to solve the system of equations and use the expansion consistently, we require the scalar field derivative $\dot \phi$ to also vary slowly compared to $f$. Therefore, we impose $\vert \ddot{\phi}L_{\rm eff}/D\vert \ll\vert \dot{\phi}\vert$. After straightforward algebra, this condition becomes
\begin{align}
   \dfrac{W''(\phi)}{W(\phi)}\ll \dfrac{D}{2(D-1)}~~.
    \label{eqn:slow_roll_eta_condition}
\end{align}
As we show below, this condition is also satisfied in the region of interest. In the dual gauge theory, it implies that the speed of sound at the critical temperature is far from its conformal value, or equivalently, that the theory is far from conformality. These are precisely the theories that are the focus of the present work.

The conditions on $W''/W$ and $(W'/W)^2$ are identical to those obtained for the Hubble slow-roll parameters $\eta$ and $\epsilon$ respectively in the context of slow-roll inflation. Inspired by this, we will use the notation $\eta \equiv W''/W$ and $\epsilon=(W'/W)^2$ throughout the paper. These conditions naturally provide a controlled setup for the adiabatic approximation \cite{Gubser:2008ny}.
\begin{figure}[t!]
    \centering
    \includegraphics[width=\linewidth]{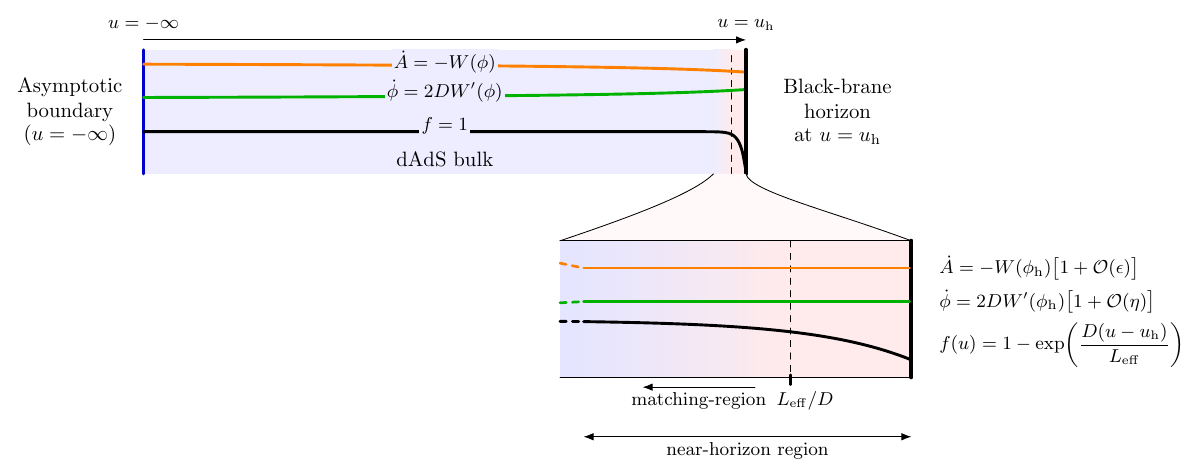}
    \caption{A schematic description of the black brane localisation. The black brane effects are localised in the region of $\mathcal{O}( L_{\rm eff}/D)$, much smaller than $L_{\rm eff}$, the local curvature scale determined by the superpotential. The bulk geometry of the dAdS spacetime can be determined using the superpotential flow equations at the leading order in $1/D$ expansion, which match onto the AdS geometry at $u \rightarrow -\infty$ and the boundary layer at $u = u_{\rm h}$.}
    \label{fig:schematic_diagram_of_large_d}
\end{figure}

\subsection{The Black Brane Solution}
We now proceed to describe the black brane solution in this approximation.
In the far-horizon region, $f(u)\simeq 1$, and $\dot{A}$ and $\dot{\phi}$ are determined by the superpotential flow equations. The boundary condition at the horizon is $f(\uh)=0$, which triggers the formation of a boundary layer of thickness $L_{\rm eff}/D$ near the horizon to correctly match the far horizon asymptotics. We show that within the boundary layer, $\dot{A}$ and $\dot{\phi}$ are constant at leading order, while the blackening function $f(u)$ varies rapidly. 

It is useful to introduce a ``stretched'' variable $z = (u-\uh)D \alpha_A$ to study the boundary layer, where $\alpha_A = \dot{A}(\uh)$. Leading corrections to this approximation are governed by $\max\{1/D,\epsilon,\eta\}$, which we denote $\zeta$. The near-horizon region is $z \ll 1/\zeta$ whereas the far region is $z \gg 1$, with an overlapping matching region $1 \ll z \ll 1/\zeta$. To analyse the equations in the near-horizon region, we employ the power counting $\dot{\phi}\sim DW'(\phih)$ and $\dot{A}\sim W(\phih)$ inspired by the far horizon solution. We will see that this power counting is self-consistent with the final result. The equation for $A$ gives 
\begin{equation}
    \dfrac{d^2 A}{dz^2}=\mathcal{O}\left(\epsilon\frac{dA}{dz}\right) \,
    \Rightarrow A(z)=A_{\rm h}+\frac{z}{D}\,.
\end{equation}
For the blackening function at leading order we get
\begin{equation}
    f(z)=\alpha_{f}(1-e^{-z})\,.
\end{equation}
The scalar field equation of motion becomes
\begin{equation}
\alpha_{A}^{2}\alpha_{f}\left(\dfrac{d^{2}\phi}{dz^{2}}(1-e^{-z})+\dfrac{d\phi}{dz}\right)=-2W(\phih)W'(\phih)\left(1+\mathcal{O}(\eta,\eta z,\epsilon z)\right)\,.
\end{equation}
The solution of this equation which is regular at the horizon is
\begin{equation}
    \phi(z)=\phih-\dfrac{2}{\alpha_{A}^2\alpha_{f}}W(\phih)W'(\phih)z\,.
\end{equation}
The solution is determined by matching the near- and far-horizon solutions in the matching region, thereby fixing the constants. In the matching region, $1/\zeta \gg z \gg 1$, the near-horizon solution reduces to
\begin{align}
    f&=\alpha_f\,,\\
    \phi(z)&=\phih-\dfrac{2}{\alpha_{A}^2\alpha_{f}}W(\phih)W'(\phih)z\,,\nonumber\\
    A(z)&=A_{\rm h}+\frac{1}{D}z\,.\nonumber
\end{align}
The far-horizon solution is the same as the vacuum solution in equation~\eqref{eqn:vacuum_superpotential} at leading order. In the matching region, it can be expanded as
\begin{align}
    f&=1\,, \\
    \phi(z)&=\phi_0+2\frac{W'(\phi_0)}{\alpha_A}z\,,\nonumber\\
    A(z)&=A_0-\frac{W(\phi_0)}{D\alpha_A} z\,.\nonumber
\end{align}
Matching gives $\alpha_f=1$, $\phi_0=\phih$, $A_0=A_h$, $\alpha_A=-W(\phih)$. Notice that this solution also satisfies the Hamiltonian constraint and the regularity condition at the horizon. Finally, we can write the leading-order solution, valid throughout the spacetime, as follows:
\begin{align}
     &\dot{A}=-W, 
     \quad \dot{\phi}=2D\dfrac{dW}{d\phi},
     \quad f(u)=1-\exp\left(D\frac{(u-\uh)}{L_{\rm eff}}\right),
     \quad L_{\rm eff}=-\dfrac{1}{\dot{A}(\uh)}\,.
     \label{eq:leadingordersolution}
\end{align}
The leading-order solution is just the vacuum solution for $A$ and $\phi$ and the effect of the black brane appears only in $f$, which ends the Euclidean geometry at the localised horizon at $u = \uh$.

In principle, one can obtain higher-order corrections by systematically expanding $f(u)$, $\dot{A}(u)$, and $\dot{\phi}(u)$ in the near-horizon region and matching the resulting solution to the far-horizon solution. We will not pursue the higher-order analysis here. Our primary interest is the thermal phase transition between the black brane and dAdS geometries and its interpretation in the dual strongly coupled gauge theory. These thermodynamic quantities and dynamics follow from the leading-order solution. 

\subsection{Thermodynamics}
\label{subsec:thermodynamics}
The Hawking temperature of the black brane is defined by requiring the absence of a conical singularity in the metric,
\begin{equation}
    T=\dfrac{e^{ A(\uh)}\vert \dot{f}(\uh)\vert}{4\pi}\,.
\end{equation}
For an AdS--Schwarzschild black brane, which is the solution in the absence of the scalar field (equation \eqref{eqn:pure_ads_blackening_factor}), the temperature is
\begin{equation}
    T_{\rm AdS}(\uh)=\dfrac{D}{4\pi L}e^{-\uh/L}~~.
\end{equation}
In this case there is no minimal temperature as a function of the horizon position $\uh$. The situation is generically different in Einstein--scalar gravity theories which model confining gauge theories away from conformality.

We now derive the Hawking temperature for an Einstein--scalar AdS black brane. It has been shown~\cite{Gursoy:2008za} that in the black brane phase the value of the scalar field at the horizon $\phi(\uh)=\phih$ uniquely determines the temperature, which we will use in the rest of the paper\footnote{It is worth clarifying that when we use $\phih$ as a temperature variable, there are potentially two different notions of derivatives. E.g.~$W'(\phih)$ could correspond to the bulk derivative of $W$ evaluated at the horizon at fixed temperature or to the difference in $W$ at the horizon at different temperatures. These two derivatives agree in our approximation.}. Using the leading-order solution in equation \eqref{eq:leadingordersolution}, we can obtain an expression for the Hawking temperature of the black brane:\footnote{
Note that $\phi$ is a monotonic function of $u$ because $\dot{\phi}$ is always positive. This follows from the dual theory having a negative holographic beta function $\beta=d\phi/dA$ and $\dot{A}$ being negative in the bulk. This allows us to go to a frame with $\phi$ as the coordinate of the warped dimension.}
\begin{equation}
    T=\dfrac{D}{4\pi}e^{A(\uh)}  |\dot{A}(\uh)|=\dfrac{D}{4\pi}\exp{\left(-\dfrac{1}{2D}\int^{\phih} d\phi \dfrac{W}{W'}\right)}  |W |\,.
\label{eqn:temperature_large_d}
\end{equation}
From this expression, we can immediately see that a minimal temperature exists ($T'(\phih) =0,~T''(\phih)>0$) when the following system of equations admits a solution  
\begin{align}
    \left(\dfrac{W'(\phih)}{W(\phih)}\right)^{2}
    &=
    \frac{1}{2D}\,,
    \nonumber\\
    \label{eq:eta}
    \dfrac{W''(\phih)}{W(\phih)}&= \frac{1+\delta}{2D} \,.
\end{align}
Here we introduce the parameter $\delta>0$, which will be useful later. If $\delta=0$, we need higher derivatives of the superpotential to establish the minimum temperature. For simplicity, we omit this possibility. 

The first equation automatically guarantees $\epsdw\ll1$. For the second condition in our approximation, $\eta \ll 1$, we require $\delta \ll D$. As noted above, this condition implies that the dual gauge theory remains far from conformality at the critical temperature.

Some further insight into the range of $\delta$ can be obtained by looking at exponential superpotentials, $W=L^{-1}(1+e^{\gamma \phi})$, well studied in the holographic literature~\cite{Gubser:2000nd,Gursoy:2007cb, Gursoy:2007er, Gursoy:2008za}. In our formalism, we find that the black brane has a minimal temperature if $\gamma\equiv(1+\delta)/\sqrt{2D}>1/\sqrt{2D}$, which is consistent with the expectation from the improved holographic QCD literature. It is also known that, for this potential, solutions with good singularities do not exist if $\gamma>\sqrt{D/2(D-1)}$ \cite{Gubser:2000nd}, hence, for this class of superpotentials, we naturally have $1+\delta\le\sqrt{D}$. We will consider the exponential superpotential in detail in section \ref{sec:exponential_example} as a test of our generic result. In this section, we calculate the maximum possible supercooling for generic superpotentials that satisfy the above conditions.

We proceed to calculate the critical temperature for the confining phase transition. In the holographic setup, different thermal phases of the strongly coupled gauge theory correspond to different geometrical solutions of Euclidean Einstein equations on the gravity side with the boundary on a thermal circle. The critical temperature $\tc$ of the phase transition between the deconfined phase (the black brane) and the confined phase (the dAdS geometry) is defined as the temperature at which the free energies of these two phases are equal (formally, the free energy is UV-divergent and needs to be regulated~\cite{Witten:1998zw, Creminelli:2001th}). There is also a thermodynamic branch corresponding to a ``small black brane'' (analogous to the small black hole solution in global AdS) which is  unstable for the branch structure we assume (see appendix~\ref{appendix:small-bh-free-energy} for details). 

\begin{figure}[t!]
    \centering
\includegraphics[width=0.9\linewidth]{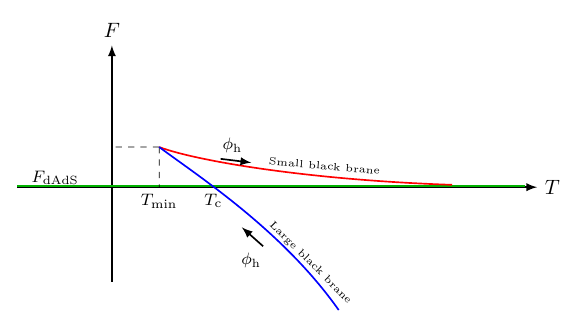}
    \caption{Schematic version of the free energy of the black brane (blue), small black brane (red) and the deformed AdS (green) solutions as a function of the temperature. The value of the field $\phih$ monotonically increases and crosses the branches at $\tmin$.}
    \label{fig:free_energy_cartoon}
\end{figure}

The generic behaviour of the free energy of these phases is shown in figure \ref{fig:free_energy_cartoon}. At high temperatures, the free energy of the black brane is the lowest, corresponding to the deconfined phase. Below $\tc$, the dAdS solution takes over, and the black brane solution becomes metastable. The free energies of the two black brane branches meet at $\tmin$, below which a black brane solution does not exist. Hence, below $\tmin$ the system has to go to thermal dAdS as this is the only thermodynamically stable solution.

The free energy difference between the two thermal solutions can be calculated using the small black brane branch \cite{Gursoy:2008za}. The black brane free energy can be related to the small black brane free energy, since they are equal at $\tmin$,
\begin{align}
    \fbh (T)
    = \fsbh(\tmin) - V_{D-1}\int_{\tmin}^T \sbb\, dT \,.
    \end{align}
The deformed AdS free energy is independent of temperature and agrees with the infinite temperature (zero mass) small black brane
    \begin{align}
    \fdads (T)
    = \fsbh(T=\infty)\,.
\end{align}
Therefore,
\begin{align}
    &\dfrac{\fbh-\fdads}{V_{D-1}}=\dfrac{\Delta F}{V_{D-1}}
    =-\int_{\infty}^{\tmin} \ssbh\, dT - \int_{\tmin}^T \sbb\, dT\,.
\end{align}
In terms of the position of the horizon parametrised by $\phih$, the integral over the two branches becomes a monotonic integral over $\phih$ (see figure~\ref{fig:free_energy_cartoon}), and the entropy density $s$ of the two branches can be written uniformly as
\begin{equation}
    s=\dfrac{e^{(D-1)A(\phih)}}{4 G_{\rm N}}=2\pi \mpld^{D-1} e^{(D-1)A(\phih)}\,.
\end{equation}
The free energy difference is therefore
\begin{align}
   \dfrac{\Delta F}{2\pi \mpld^{D-1}V_{D-1}} =-\int_{\infty}^{\phih} e^{(D-1)A(\widetilde{\phi}_{\rm h})}\dfrac{dT}{d\widetilde{\phi}_{\rm h}}~d\widetilde{\phi}_{\rm h}\,.
    \label{eqn:free_energy_difference}
\end{align}

The critical temperature corresponds to $\Delta F(\phih)=0$, whose solution we denote by $\phih=\phic$. In the $1/D$ expansion we can again obtain the critical temperature in a model-independent way. Under mild assumptions about the IR asymptotics of the theory (see appendix~\ref{appendix:small-bh-free-energy}) the factor $e^{DA}$ in the expression for the free energy localises the integral around the lower boundary of the integral within a region of width $\sim 1/(D |A'(\phimin)|)$. Therefore, the free-energy difference is zero for $\phic$ close to $\phimin$ where the integrand switches sign. We obtain
\begin{equation}
  \dfrac{\Delta F(\phic)}{2\pi \mpld^{D-1}V_{D-1}}
  =
  \dfrac{2 e^{DA(\phimin)}T(\phimin)}{D^{2}} 
  \left(\delta e^{-D\frac{\phic-\phimin}{\sqrt{2D}}}
  \left(1+D\dfrac{\phic-\phimin}{\sqrt{2D}}\right)
  +\mathcal{O}\left(\frac{1}{D} \right)\right)\,.  \label{eqn:free_energy_diffrence_expanded}
\end{equation}
Solving for $\Delta F(\phi_c)=0$ gives 
\begin{equation}
   \dfrac{\phimin-\phic}{\sqrt{2D}}=\dfrac{1}{D}~~,
\end{equation}
 We can now obtain the critical temperature using equation~\eqref{eqn:temperature_large_d} as follows:
\begin{equation}
T_{\rm c}=T_{\rm min}+\dfrac{1}{2}T''(\phic-\phimin)^2+\cdots=T_{\rm min}\left(1+\frac{\delta}{D^2}\left(1+\mathcal{O}\left(\dfrac{1}{D},\dfrac{\delta}{D}\right)\right)\right)\,,
\label{eqn:T_c,t_in_relation_in_terms_of_delta}
\end{equation}
Hence, using this relation we find the maximum possible supercooling to be
\begin{equation}
  \epsilon_{\rm sc}=1-\dfrac{T_{\rm min}}{T_{\rm c}}=\dfrac{\delta}{D^2}\left(1+\mathcal{O}\left(\dfrac{1}{D},\dfrac{\delta}{D}\right)\right)\,.
\label{eqn:maximal_supercooling_in_terms_of_delta}
\end{equation}
The expression is valid in the regime $1/D\ll\delta\ll D$. We see that the maximum supercooling is parametrically suppressed by $1/D^2$, suggesting that it is generically very small. 

It is intriguing that the same $1/D^2$ scaling also appears in the deconfinement--confinement transition of $\mathcal{N}=4$ Super Yang--Mills theory on $S^{D-1}\times S^{1}$ \cite{Witten:1998zw, Agrawal:2025wvf}. Furthermore, there are hints towards small supercooling in thermal pure $SU(N)$ Yang--Mills theory in the large-$N$ limit \cite{Lucini:2002ku,Pasechnik:2023hwv,Agrawal:2025xul, Huber:2025qbl}, consistent with the order-of-magnitude estimate obtained by extrapolating the $1/D^2$ scaling to $D=4$. Taken together, these observations may point towards a universal property of strongly coupled gauge theories. 

\subsection{Holographic interpretation}
\label{subsec:holography}
The supercooling in equation \eqref{eqn:maximal_supercooling_in_terms_of_delta} can be interpreted as the maximum possible supercooling of the deconfined phase in the dual theory. It is useful to relate this result to a physical observable in the gauge theory. For this purpose, using the speed of sound is particularly convenient, since it vanishes at the minimal temperature. We can calculate its value at the critical temperature\footnote{The sound speed at the critical temperature in a first-order transition is formally ill-defined. The sound speed here refers to the smooth continuation of the sound speed of the deconfined phase.} using a Taylor expansion as follows:
\begin{align}
    \cs^2(T_{\rm c})=\dfrac{d\log{T}}{d\log {s}}=\frac{1}{D}\dfrac{d\log{T}}{d\phi}\dfrac{d\phi}{dA}=-2\dfrac{\delta}{D}\frac{(\phic-\phimin)}{\sqrt{2D}}\left(1+\mathcal{O}\left(\dfrac{1}{D},\dfrac{\delta}{D}\right)\right)\nonumber\\=2\dfrac{\delta}{D^2}\left(1+\mathcal{O}\left(\dfrac{1}{D},\dfrac{\delta}{D}\right)\right) \,.
\end{align}
In the expression above, the ratio $\delta/D$ can be replaced by $c_{\rm s}^{2}(\tc)/\ccft^{2}$ where $\ccft^{2}=1/(D-1)$ is the speed of sound in the conformal limit in $D$ dimensions. As a result, the leading-order expression for the maximum supercooling is
\begin{align}
\boxed{
    \epsilon_{\rm sc}=\dfrac{\cs^2(T_{\rm c})}{2}
    \label{eqn:maximal_supercooling}
    }
    \,,
\end{align}
with relative corrections of $\mathcal{O}(1/D)$ and $\mathcal{O}((\cs(\tc)/{\ccft})^{2})$.
As our approximation scheme is valid in the regime $1/D\ll\delta\ll D$, in terms of the speed of sound, it translates to 
\begin{equation}
    \dfrac{1}{D^{2}}\ll \left(\dfrac{\cs(\tc)}{\ccft}\right)^{2}\ll 1~~.
\end{equation}

Note that the ratio $(\cs(\tc)/\ccft)^{2}$ is a measure of deviation from conformality at $\tc$, meaning that $\cs^{2}(\tc)$ far from $\ccft^{2}$, implies being away from conformality. As $\cs(T_{\rm c})$ approaches zero, our leading-order functional form is no longer valid. Nevertheless, in this limit, by definition, $T_{\rm c}$ approaches $T_{\rm min}$ and supercooling goes to zero. However, the precise functional form of this approach cannot be determined without additional information about the superpotential. 

As $\cs^{2}(\tc)$ increases towards $\ccft^{2}$ the supercooling increases. When the theory moves closer to conformality at $\tc$, the expansion parameter $(c_{\rm s}(\tc)/\ccft)^{2}~\sim \mathcal{O}(1)$, the approximation breaks down and the maximum supercooling cannot be calculated using our formula. In this limit, the near-conformal theories have been studied using effective dilaton models \cite{Creminelli:2001th, Randall:2006py, Konstandin:2011dr, Baratella:2018pxi, Agashe:2019lhy, Agashe:2020lfz} which show that the supercooling can be $\mathcal{O}(1)$.
\section{Comparison with explicit examples}
\label{sec:exponential_example}
\begin{figure}[t!]
    \centering
    \includegraphics[width=0.45\linewidth]{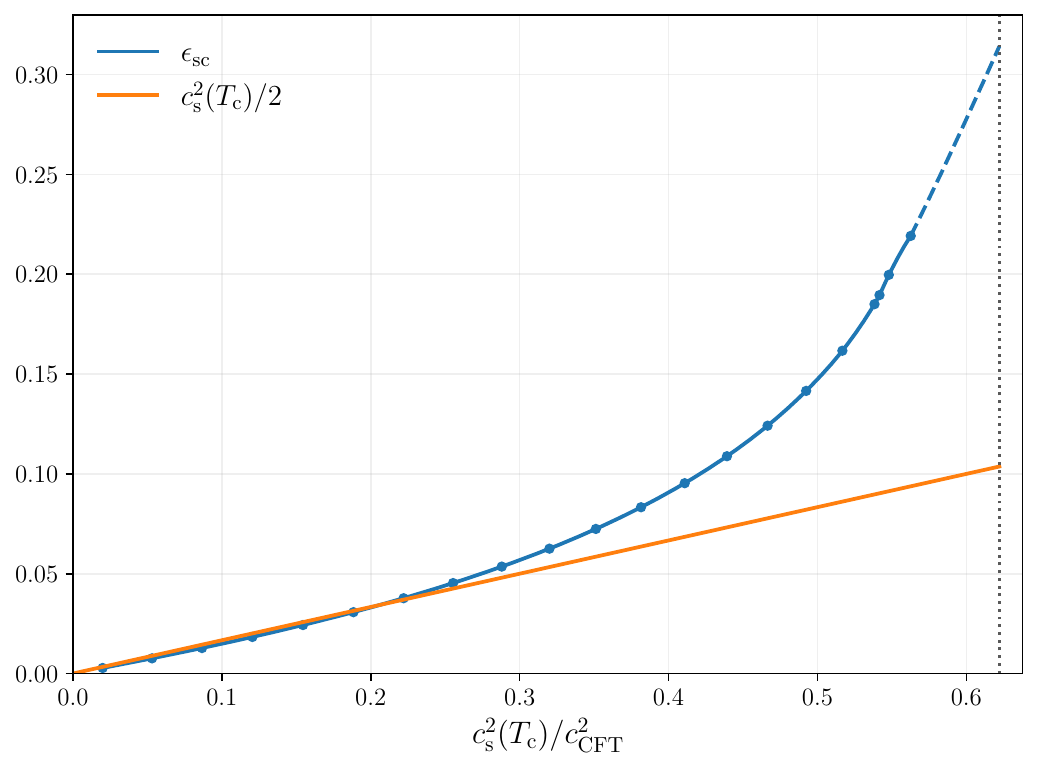}
    \includegraphics[width=0.45\linewidth]{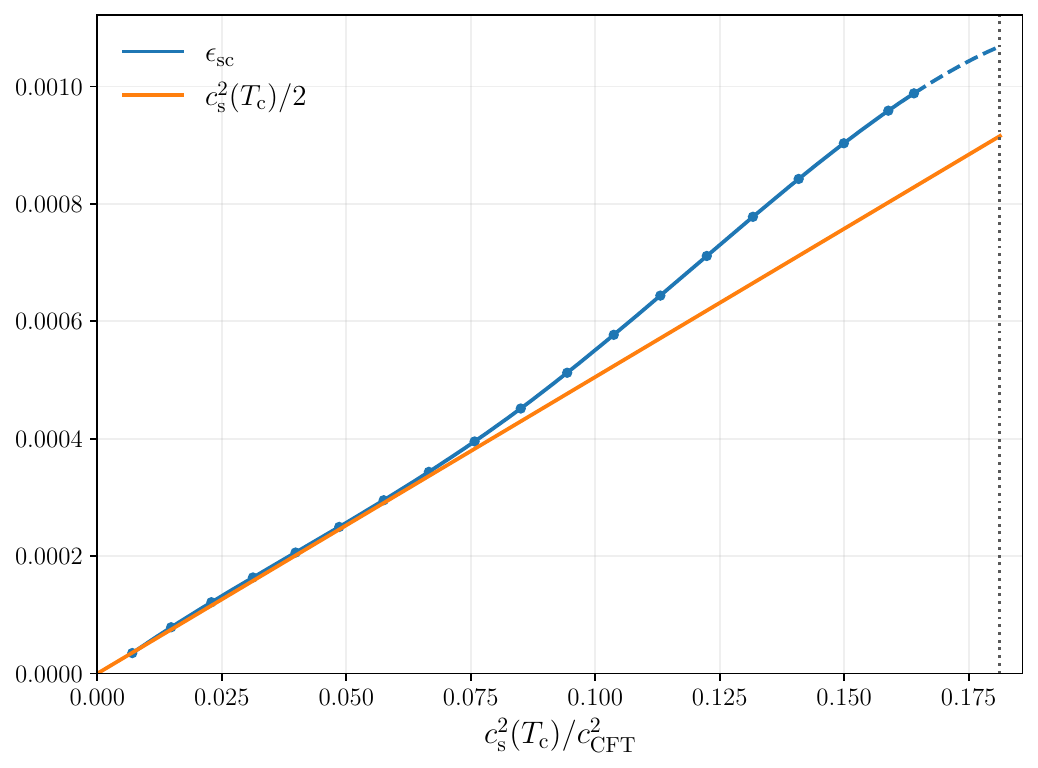}
    \caption{The maximum supercooling $\epsilon_{\rm sc} = 1-T_{\rm min}/\tc$ as a function of the ratio of squared sound speeds $(\cs(\tc)/\ccft)^{2}$ for $D=4$ (\textbf{left panel}) and $D=100$ (\textbf{right panel}). We see that numerical results obtained for the potential \eqref{eqn:exponential_superpotential} agree very well with our general result \eqref{eqn:maximal_supercooling} in the region where $(\cs(\tc)/\ccft)^{2}$ is small. The agreement becomes better when the number of dimensions increases.}
\label{fig:supercooling_vs_soundspeed}
\end{figure}
In this section, we compare our general prediction for maximum supercooling, equation~\eqref{eqn:maximal_supercooling}, with explicit examples, namely the exponential superpotential, improved holography models and $\mathcal{N}=4$ SYM, and find good agreement. 
\paragraph{The exponential superpotential:}
We compare our general prediction for the maximum  supercooling, equation~\eqref{eqn:maximal_supercooling}, with the direct numerical results obtained in Einstein--scalar gravity for the following superpotential:
\begin{equation}
    W=\dfrac{1}{L}(1+e^{\gamma \phi})\,,
\label{eqn:exponential_superpotential}
\end{equation}
which has been extensively studied in the context of improved holography \cite{Gursoy:2007cb, Gursoy:2007er} and the Hawking--Page-type transition \cite{Mishra:2024ehr, Ismail:2026tyb}. In the absence of the black brane, the solution for the warp factor and the scalar field profile are
\begin{align}
     &   A(\phi)=-\dfrac{1}{2(D-1)\gamma}
     \left(\phi-\dfrac{1}{\gamma}e^{-\gamma\phi}\right),\nonumber\\
     &   \phi(u)=-\dfrac{1}{\gamma}\ln\left(-\dfrac{2(D-1)\gamma^{2}(u-u_{\star})}{L}\right)\,.
\end{align}
    where $u_{\star}$ is the location of the IR singularity. 
In this convention, we can see that as one approaches the asymptotic boundary $u\rightarrow -\infty$ the warp factor $A(u)\rightarrow -u/L$ up to logarithmic corrections indicating asymptotic $\text{AdS}$ behaviour. For the superpotential of the form in equation \eqref{eqn:exponential_superpotential} $\gamma$ has to lie between
\begin{equation}
\dfrac{1}{\sqrt{2(D-1)}}\leq \gamma \leq \sqrt{\dfrac{D}{2(D-1)}}~~.
\end{equation}
The lower bound makes sure that the dual theory has a confining vacuum \cite{Gursoy:2008za} whereas the upper bound comes from the well-behaved nature of the IR singularity according to the Gubser criterion \cite{Gubser:2000nd}. At the upper end of the Gubser criterion, the squared speed of sound is approximately equal to $c_s^2 \simeq 2\ccft^2/\sqrt{D}$ at leading order in $1/D$.

As a black brane solution cannot be obtained analytically, we numerically constructed it and evaluated both the maximum supercooling and the speed of sound in the deconfined phase at the critical temperature. The comparison with the analytical prediction, equation~\eqref{eqn:maximal_supercooling}, is shown in figure~\ref{fig:supercooling_vs_soundspeed}.\footnote{The initial version of this figure was successfully generated by ChatGPT 5.6 using the rather general prompt: {\tt Can you take Einstein-Scalar gravity in 5D with superpotential $W=L^{-1}(1+e^{\gamma\phi})$ and numerically find solution of ratio of minimal and critical temperatures associated with black hole to vacuum phase transition as a function of $\gamma$}. This unexpected success led one of the authors (VL) to abandon his previously held mildly sceptical view of AI.} The figure shows that the agreement between the general analytical prediction and the numerical results obtained for the specific superpotential is very good in the regime where the theory is far from conformality. As expected, the agreement improves as the number of spacetime dimensions increases. This is partly because the $1/D$ corrections become smaller and partly because, in this model, the speed of sound decreases throughout the range as $D$ increases. We see that the agreement is very good for the physically relevant case of $D=4$ in the region where $(\cs(\tc)/\ccft)^2\lesssim0.4$. Beyond this point, our result should be regarded as an order-of-magnitude estimate, as the approximation is pushed towards the edge of its regime of validity.

\paragraph{Improved holography:} We can compare our prediction with specific calculations in the improved holography model studied in~\cite{Gursoy:2010fj, Missoni:2026mvx}. The sound speed in this model calculated in~\cite{Gursoy:2010fj} is $\cs^2(\tc)\simeq 0.11$, and the maximum supercooling in~\cite{Missoni:2026mvx}, $\epsilon_{\rm sc} \simeq 0.04$, showing encouraging agreement with our estimate. 

\paragraph{$\mathcal{N}=4$ SYM:} We can compute the supercooling in the confinement transition in $\mathcal{N}=4$ SYM on $S^3\times S^1$ generalised to $D$ dimensions. On the spatial sphere, we do not have the same notion of sound speed, but to uniformly compare with our results we define $\cs$ as the same thermodynamic quantity $\cs^2 = d\log T/d \log s$, which evaluated at the critical temperature gives $\cs^{2}(\tc)=1/(D-1)^{2}$. Then we find that the supercooling is,
\begin{align}
\epsilon_{\rm sc} 
=
1-\sqrt{1-\cs^2(\tc)}\, ,
\end{align}
which strikingly agrees with our estimate for $\cs^2(\tc) \ll 1$.

\paragraph{The lattice:} The deconfined phase sound speed has been calculated on the lattice for $SU(3)$ Yang-Mills close to $\tc$, $\cs^2(\tc) \simeq 0.013$~\cite{Giusti:2025fxu}, suggesting sub-percent supercooling in $SU(3)$. It will be nice to have a direct lattice check of these predictions.

\section{Conclusion}
We study confinement transitions in strongly coupled gauge theories using Einstein--scalar holography, and solve the system, in a $1/D$ expansion for a large class of scalar potentials far from conformality. We provide a prediction for the maximum possible supercooling for the metastable deconfined phase below the critical temperature.  Besides its intrinsic importance for the dynamics of confinement, this question is directly relevant for cosmological phase transitions, where the amount of supercooling dramatically affects the strength of the resulting gravitational-wave signal.

We use the superpotential formalism augmented by the $1/D$ expansion, which allows us to solve the black brane geometry for a large class of scalar potentials. The blackening function becomes localised near the horizon over a distance scale of order $L_{\rm eff}/D$, while the remaining bulk geometry varies on the larger local curvature scale $L_{\rm eff}$. This allows us to match the near-horizon region to a far-horizon region where the scalar field and warp factor are governed by the superpotential flow equations. 

Using these solutions, we analyse the Hawking--Page transition between the black brane and dAdS geometries, which is interpreted holographically as a thermal deconfinement--confinement transition in the dual strongly coupled gauge theory. We show that we have control in a broad class of theories far from conformality, in which the maximum supercooling is parametrically suppressed as
\begin{equation*}
\epsilon_{\rm sc}\sim \frac{1}{D^2}\,.
\end{equation*}
The maximum supercooling can be related directly to the speed of sound in the deconfined phase at the critical temperature. At leading order in the $1/D$ expansion, we find
\begin{equation*}
\epsilon_{\rm sc}=\frac{\cs^2(T_{\rm c})}{2}\,.
\end{equation*}
This relation gives a useful physical interpretation of the result: theories that have a small speed of sound near the transition are far from conformality and have a small maximum supercooling. For these theories, $\cs^2\ll 1/3$, the relation implies the bound $\epsilon_{\rm sc}\ll 1/6$ within our approximation. We emphasise that our analysis does not apply to theories with approximate conformal invariance, where the assumptions underlying the expansion break down and large supercooling can occur.

We have compared our prediction with explicit results for an exponential superpotential, improved holography and $\mathcal{N}=4$ SYM on a sphere. Together with indirect evidence for small supercooling in YM~\cite{Agrawal:2025xul}, quantitatively consistent with the magnitude expected from a $1/D^2$ scaling, this suggests that the behaviour may be a rather general feature of strongly coupled theories far from conformality. It will be interesting to understand to what extent this finding is universal. The lattice determination of the sound speed in $SU(3)$ YM theory suggests that the theory may lie within the validity of our approximation, and it would be extremely interesting to explicitly calculate supercooling in this theory on the lattice. 

\subsubsection*{Acknowledgements}
We thank Roberto Emparan, Juan Maldacena, John March-Russell, Georges Obied, and Diego Redigolo for helpful discussions. We acknowledge the use of OpenAI Codex for the purposes of code development and polishing prose. All the results are analysed and verified by the authors. The work of PA was supported in part by the U.S. Department of Energy under the grant DE-SC0011702. This work was performed in part at the Aspen Center for Physics, which is supported by National Science Foundation grant PHY-2210452. GRK expresses gratitude for support via the Somerville College Oxford Ryniker Lloyd Graduate Scholarship jointly with a Clarendon Fund Scholarship. VL is supported by the STFC grant ST/X000761/1. For the purpose of Open Access, the authors have applied a CC BY public copyright licence to any Author Accepted Manuscript version arising from this submission.
\appendix 
\section{Details of the equations of motion}
\label{appendix:details_of_the_equations_of_motion}
Consider the metric with the domain wall ansatz given in equation \eqref{eqn:warp_metric_ansatz} for a $(D+1)$-dimensional spacetime. In this appendix, we derive the equations of motion.  One can verify that for the ansatz in equation \eqref{eqn:warp_metric_ansatz} the Ricci tensor is diagonal. The non-vanishing components of the Ricci tensor are:
\begin{align}
  &  R_{tt}=-f e^{2A}\left(\dfrac{1}{2}\ddot{f}+\ddot{A}f+D\dot{A}^{2}f+\dfrac{D+2}{2}\dot{A}\dot{f}\right)\nonumber\\
&    R_{uu}=-\left(D\ddot{A}+D\dot{A}^{2}+\dfrac{D+2}{2}\dfrac{\dot{A}\dot{f}}{f}+\dfrac{\ddot{f}}{2 f}\right)\\
&R_{ij}=-\delta_{ij}e^{2A}\left(\dot{f}\dot{A}+\ddot{A}f+Df\dot{A}^{2}\right)\nonumber
\end{align}
With the action in equation \eqref{eqn:einstein_scalar_action}, the Einstein field equations can be re-expressed as:
\begin{equation*}
    R_{ab}=\dfrac{1}{2}\partial_{a}\phi\partial_{b}\phi+\dfrac{g_{ab}}{D-1}V(\phi)~~.
\end{equation*}
Substituting the various components of the Ricci tensor, one finds:
\begin{align}
   & R_{tt}=g_{tt}\dfrac{V}{D-1}\Rightarrow \dfrac{1}{2}\ddot{f}+\ddot{A}f+D\dot{A}^{2}f+\dfrac{D+2}{2}\dot{A}\dot{f}=-\dfrac{V}{D-1}\nonumber\\
 &   R_{uu}=\dfrac{1}{2}\dot{\phi}^{2}+\dfrac{g_{uu}}{D-1}V\Rightarrow D\ddot{A}+D\dot{A}^{2}+\dfrac{D+2}{2}\dfrac{\dot{A}\dot{f}}{f}+\dfrac{\ddot{f}}{2 f}=-\dfrac{1}{2}\dot{\phi}^{2}-\dfrac{V}{f(D-1)}
    \label{eqn:ricci_tensor_eom}\\
&    R_{ij}=g_{ij}\dfrac{V}{D-1}\Rightarrow \dot{f}\dot{A}+\ddot{A}f+Df\dot{A}^{2}=-\dfrac{V}{D-1}\nonumber
\end{align}
The first two equations associated with $R_{tt}$ and $R_{uu}$ can be rearranged as
\begin{equation}
    \ddot{A}=-\dfrac{1}{2(D-1)}\dot{\phi}^{2}~~.
\end{equation}
Substituting this relation between $\dot{\phi}$ and $\ddot{A}$ in the equation associated with $R_{ij}$ in equation \eqref{eqn:ricci_tensor_eom} gives 
\begin{align}
    D(D-1)f\dot{A}^{2}+(D-1)\dot{A}\dot{f}=\dfrac{f\dot{\phi}^{2}}{2}-V~~.
\end{align}
Finally, we have the scalar field equation of motion given by
\begin{align}
&\dfrac{1}{\sqrt{g}}\partial_{\mu}(\sqrt{g}g^{\mu\nu}\partial_{\nu}\phi)=\dfrac{dV}{d\phi}
\Rightarrow \dot{f}\dot{\phi}+f\ddot{\phi}+Df\dot{A}\dot{\phi}=\dfrac{dV}{d\phi}~~.
\end{align}
The second equality is implied by the fact that $\phi$ only depends on the $u$ coordinate. 
To summarise all the equations, we have
\begin{align}
 &\ddot{A}=-\dfrac{1}{2(D-1)}\dot{\phi}^{2}~,~~~
D(D-1)f\dot{A}^{2}+(D-1)\dot{A}\dot{f}=\dfrac{f\dot{\phi}^{2}}{2}-V\nonumber\\
 & \dfrac{\ddot{f}}{\dot{f}}+D\dot{A}=0~,~~
\dot{f}\dot{\phi}+f\ddot{\phi}+Df\dot{A}\dot{\phi}=\dfrac{dV}{d\phi}~~.
\end{align}

\section{Free energy of the small black brane branch}
\label{appendix:small-bh-free-energy}
In this appendix, we derive equation \eqref{eqn:free_energy_diffrence_expanded} using the $1/D$ expansion and state the conditions under which the integral is dominated near $\phimin$. We start by
splitting the integral in equation \eqref{eqn:free_energy_difference} at $\phi=\phi_\star$, 
\begin{equation}
\int_{\infty}^{\phic} e^{(D-1)A(\phih)}\dfrac{dT}{d\phih}~d\phih
=   
\underbrace{\int_{\phi_{\star}}^{\phic} e^{(D-1)A(\phih)}\dfrac{dT}{d\phih}~d\phih}_{\mathcal{I}_{1}}
+\underbrace{\int_{\infty}^{\phi_{\star}} e^{(D-1)A(\phih)}\dfrac{dT}{d\phih}~d\phih}_{\mathcal{I}_{2}}~~.
\label{eqn:free_energy_integral_break_down}
\end{equation}
We choose $\phi_\star$ such that
\begin{equation}
    \dfrac{1}{D}\ll \dfrac{\phi_{\star}-\phimin}{\sqrt{2D}}\ll \dfrac{1}{\delta}~~.
\label{eqn:choice_of_phi_star}
\end{equation}
Since $\delta\ll D$, this choice is always possible. In this case, the first integral in equation \eqref{eqn:free_energy_integral_break_down} can be evaluated using the Taylor series expansion of $A(\phih)$ and $dT/d\phih$ around $\phih=\phimin$:
\begin{align}
A(\phih) &=A(\phimin)+A'(\phimin)(\phih-\phimin)+\dfrac{1}{2}A''(\phimin)(\phih-\phimin)^{2}+\cdots\\
   \dfrac{dT}{d\phih}
   &=\dfrac{dT}{d\phih}(\phimin)+\dfrac{d^{2}T}{d\phih^{2}}(\phimin)(\phih-\phimin)+\cdots
\end{align}
In this region, the slow-roll conditions are satisfied, allowing us to use our approximation, in which we can express $A$ and its derivatives in terms of the superpotential value at $\phih=\phimin$
\begin{equation}
A'(\phimin)=-\dfrac{1}{\sqrt{2D}}~,~
    A''(\phimin)=\dfrac{\delta}{2D}.
\end{equation}
Similarly, at $\phih=\phimin$, $dT/d\phih=0$ and the second derivative takes the form
\begin{align}
    \dfrac{d^{2}T}{d\phih^{2}}\Bigg\vert_{\phih=\phimin}=\dfrac{\delta}{D}\tmin~~.
\end{align}
We note that when the condition in equation \eqref{eqn:choice_of_phi_star} is satisfied and $1/D\ll \delta \ll D$, it is sufficient to keep only the linear terms in the Taylor series expansion to evaluate the integral. Substituting this into the expression for $\mathcal{I}_{1}$ gives
\begin{align}
 \mathcal{I}_{1}
 &=
 2\delta e^{(D-1)A(\phimin)}T(\phimin)
 \int_{(\phi_{\star}-\phimin)/\sqrt{2D}}^{(\phic-\phimin)/\sqrt{2D}} e^{-(D-1)x}~x~dx
 \\\nonumber
 &=
 -
 \left.
 \frac{2\delta e^{(D-1)A(\phimin)}T(\phimin)}{(D-1)^2}
 \left[
 e^{-(D-1)x} (1+(D-1)x)
 \right]
 \right|_{(\phi_{\star}-\phimin)/\sqrt{2D}}^{(\phic-\phimin)/\sqrt{2D}} 
\label{eqn:free_energy_difference_large_d}
\end{align}

According to the choice of $\phi_{\star}$ in equation \eqref{eqn:choice_of_phi_star}, the contribution from the lower limit of the integral is exponentially suppressed and hence can be ignored in the large-$D$ limit.

Let us now look at the second integral in equation \eqref{eqn:free_energy_integral_break_down}. In the region where the slow-roll conditions are satisfied, we know that the value of $A$ at the horizon is a monotonically decreasing function of $\phih$. In the far-IR asymptotic region, it is not always guaranteed that the slow-roll conditions will be satisfied; hence, we cannot directly relate $A$ to the vacuum $\beta$ function $W'/W$. However, as noted in \cite{Gursoy:2008za}, this condition is typically satisfied in the deep IR region and can only be violated in a small finite region. The violation of the condition corresponds to the small black brane branch becoming thermodynamically stable or $T'(\phih)$ switching sign, which would generate additional local extrema in the temperature. Here, we assume a minimal case in which such a situation is excluded. Moreover, even if this happens, as we have already made sure that the warp factor $A$ sufficiently decreases before reaching this region where the condition can be violated, the corrections from this at most finite region should be small, although the exact quantitative value of this correction would need the full functional form of the superpotential.  

To estimate the second integral, we assume for simplicity that $A(\phih)$ is a monotonically decreasing function going to $-\infty$ at the IR singularity such that $dA(\phih)/d\phih$ asymptotes to a finite non-zero value \cite{Gursoy:2008za} and hence $A(\phih)<A(\phi_{\star})-c(\phih-\phi_{\star})$ where $c=\inf_{\phih>\phi_{\star}}\vert dA(\phih)/d\phih\vert>0$. This allows us to bound the magnitude of the integral $|\mathcal{I}_{2}|$ from above using the pointwise inequality of the exponential function in the integrand giving 
\begin{align}
    |\mathcal{I}_{2}|
    \leq 
    e^{(D-1)A(\phimin)}e^{-(D-1)\frac{\phi_{\star}-\phimin}{\sqrt{2D}}}
    \left|
    \int_{\infty}^{\phi_{\star}} e^{-(D-1) c(\phi-\phi_{\star})}\dfrac{dT}{d\phi}~~d\phi
    \right|
    \,.
\end{align}
After using the  weighted mean value theorem, the upper bound on $\mathcal{I}_{2}$ magnitude can be written as
\begin{equation}
    |\mathcal{I}_{2}|
    \leq 
    e^{(D-1)A(\phimin)}e^{-(D-1)\frac{\phi_{\star}-\phimin}{\sqrt{2D}}}
    \dfrac{1}{(D-1)c}\left|\dfrac{dT}{d\phi} (\phi_{0})\,\right|,
\end{equation}
where $\phi_{0}\in [\phi_{\star},\infty)$. Therefore, in the limit $D\rightarrow \infty$,  $\mathcal{I}_{2}$ is exponentially suppressed compared to $\mathcal{I}_{1}$ as long as $c$ is not exponentially small or $\tfrac{dT}{d\phi}(\phi_0)$ is not exponentially large, both of which are very mild conditions on the IR asymptotics. Notice that this exponential suppression is generated in the region where we have full control -- the asymptotic integral determines the coefficient of the exponential and as seen above, it does not compensate the suppression. 

We can explicitly check the calculation of $\mathcal{I}_2$ in the case of a pure exponential superpotential $W(\phi) = e^{\gamma \phi}$. In this case, we have an analytical solution for $A(\phi), T(\phi)$, 
\begin{align}
    \mathcal{I}_2
    &=
    -\frac{2(D-1)\gamma^2-1}{D-2(D-1)\gamma^2}
    T(\phi_\star)
    e^{(D-1) A(\phi_\star) }
    \\ \nonumber
    &\simeq
    -\frac{2(D-1)\gamma^2-1}{D-2(D-1)\gamma^2}
    \tmin
    e^{(D-1)A(\phimin)}e^{-(D-1)\frac{\phi_{\star}-\phimin}{\sqrt{2D}}}\,.
\end{align}
We can see explicitly the exponential suppression in this integral for all $\gamma$ in the allowed Gubser region $\gamma < \sqrt{D/(2(D-1))}$. To summarise, at leading order in the $1/D$ expansion, the free energy difference is given by
\begin{equation}
  \dfrac{\Delta F(\phic)}{2\pi \mpld^{D-1}V_{D-1}}=\dfrac{2\delta e^{DA(\phimin)}T(\phimin)}{D^{2}}\times \left(e^{-D\frac{\phic-\phimin}{\sqrt{2D}}}\left(1+D\dfrac{\phic-\phimin}{\sqrt{2D}}\right)\right)\,.
\end{equation}
  \bibliography{references}
\bibliographystyle{JHEP}  
\end{document}